\documentclass[apj]{emulateapj}

\usepackage{amssymb, aas_macros}
\usepackage{color}
\usepackage{natbib}
\newcommand{\MSF}{{{\rm M}_{\rm SF}}}
\newcommand{\Mgas}{{{\rm M}_{\rm gas}}}
\newcommand{\Mstar}{{{\rm M}_{\rm star}}}
\newcommand{\Mvir}{{{\rm M}_{200}}}
\newcommand{\Rvir}{{{\rm R}_{200}}}

\newcommand{\denv}{{\delta_{1200}}}
\newcommand{\lcdm}{{$\Lambda$CDM}}
\newcommand{\SCNinprep}{{Scannapieco et al., submitted}}
\newcommand{\SCNbinprep}{{Scannapieco et al. (submitted)}}
\begin{document}

\title{The Effect of Environment on Milky Way-mass galaxies in a Constrained Simulation of the Local Group}

\author{Peter Creasey\altaffilmark{1}}
\author{Cecilia Scannapieco\altaffilmark{1}}
\author{Sebasti\'an E. Nuza\altaffilmark{1}}
\author{Gustavo Yepes\altaffilmark{2}}
\author{Stefan Gottl\"ober\altaffilmark{1}}
\author{Matthias Steinmetz\altaffilmark{1}}

\altaffiltext{1}{Leibniz-Institut f\"ur Astrophysik Potsdam (AIP), An der Sternwarte 16, D-14482, Potsdam, Germany}
\altaffiltext{2}{Grupo de Astrof\'{\i}sica, Universidad Aut\'onoma de Madrid, Madrid E-28049, Spain}

\begin{abstract}
In this letter we present, for the first time, a study of star formation rate, gas fraction and galaxy morphology of a constrained simulation of the Milky Way (MW) and Andromeda (M31) galaxies, compared to other MW-mass galaxies.
By combining with unconstrained simulations we cover a sufficient volume to compare these galaxies environmental densities ranging from the field to that of the Local Group (LG). This is particularly relevant as it has been shown that, quite generally, galaxy properties depend intimately upon their environment, most prominently when galaxies in clusters are compared to those in the field.
For galaxies in loose groups such as the LG, however, environmental effects
have been less clear.
We consider the galaxy's environmental density in spheres
of 1200 kpc (comoving)  and find that whilst environment
does not appear to  directly affect morphology, there is
a positive trend with star formation rates. This enhancement
in star formation occurs systematically for galaxies in higher density environments, regardless whether they are part of the
LG or in filaments.
Our simulations suggest that the richer environment at Mpc-scales may help replenish the star-forming gas, allowing higher specific star formation rates in galaxies such as the MW.

\end{abstract}

\maketitle

\section{Introduction}

During the last decades, 
observational studies
found indications that the  environment where galaxies
form  plays an important role on the determination
of their final properties. For example, 
elliptical galaxies are more strongly clustered than
spirals \citep[e.g. ][]{Dressler80, Hermit96, Guzzo97}; and correlations 
between environment and the photometric colour or luminosity of a galaxy
have also been measured \citep{Blanton03}. 
The suppression of star formation in clustered environments also extends to groups \citep{Lewis_2002, Coenda_2014} and the morphology trends, i.e. that early type galaxies lie closer to halo centres, extend as a continuum to group scales \citep{Girardi_2003}, although at stellar masses closer to that of the MW this is more controversial \citep{Ziparo_2013} and some authors suggest large scale features such as filaments may be more important than the group \citep[e.g. ][]{Bahe_2013}.
Group galaxies also appear deficient in cold gas \citep{Hess_2013}, although \citet{McGee_2008} find that whilst the disc fraction depends on the group environment, the other properties of the discs appear largely unaffected. 

On the other hand, theoretical studies in the context of $\Lambda$-Cold Dark Matter (\lcdm)  show that
the merger rate of galaxies depends
on environment, such that galaxies in intermediate and high-density
regions have the  highest fractions of mergers (e.g. \citealp{Maulbetsch_2007,deRavel_2009, Darg_2010, Ellison_2010, Fakhouri_2010, Lin_2010,Tonnesen_2012, Jian_2012}).
Mergers are indeed known to induce  morphological  transformations
of galaxies \citep[e.g. ][]{Toomre_1972}, and could be responsible for at least some of the
observed trends of galaxy properties with environment.

Our MW lives in a rich group environment, 
known as the ``Local Group" (LG).
Andromeda (M31), the other large spiral in the LG, has a similar mass to the MW, 
and lies at less than a Mpc away.
Such spiral galaxies are prevalent in the Universe, and thus a fundamental test
of the 
\lcdm\ cosmological framework is that neither the local universe \citep[e.g. ][]{Nuza_14b}
nor the MW and M31 should be extremely improbable objects.

The numerical simulation of MW-like galaxies is an active area of research, with
zoom-in techniques \citep{Katz_1993} allowing the collisionless and collisional
components of cosmological density fluctuations to be followed over 
almost the full $13.7$ billion years of cosmic time.
Modern computers enable simulations that describe
the internal properties of galaxies so that time variations
due to mergers, interactions and infall can indeed be quantified
using an acceptable
number of resolution elements.

The precise 
initial conditions (ICs) to produce late type galaxies such as the MW and M31
are not believed to be
strict as spiral galaxies are the most abundant type in our Universe. In 
most modern studies \citep[e.g. ][]{S09, Guedes_2011, Aumer13,Stinson_2013, Vogelsberger_2014}
the ICs are chosen as halos
of MW mass
and isolated from other massive halos that can
destroy discs at late times. 
Indeed the primary uncertainty of such  
simulations lies in the unresolved 
baryonic physics that result in star formation and feedback 
\citep[see][]{Creasey_2011}, and different codes 
applied to the same halo can produce substantial variation in the stellar
component \citep{S12}.

Whilst the interaction of the M31-MW pair is not yet expected to affect internal
galaxy properties as they are still too distant, their environments on Mpc scales
have merged, and this overdense environment may affect the pair.
One method to test such effects is to identify 
analogous pairs produced by cosmological ICs, and compare 
to  more isolated systems.
Following this approach, 
\citet{Few12}  found that 
for simulated MW-mass galaxies the difference between the field and 
loose groups appears marginal, and in fact they detected no
visible difference between MW-like field galaxies and those that
reside in loose groups. 
\citet{GarrisonKimmel_2014} find
no difference in concentrations or stellar masses within the virial 
radius, but do find an increased number of `backsplash' galaxies 
\citep[see e.g.][]{Knebe_2011} - those
that have escaped the virial radius - for pairs such as the MW and M31.

Although those simulations allow assessment of the effects of environment
on the formation of MW-mass galaxies, they do not fully
exploit the detailed constraints
we have from the present-day dynamics of our local universe.
One method to ameliorate this is to utilise the 
power of  {\it constrained} ICs
that reproduce the observed dynamical properties of the LG
in combination with hydrodynamical simulations, which has not 
previously been attempted.
This provides a pertinent test to quantify the influence of environment,
as well as allowing us to 
discern directly the physical processes in play during the
formation of galaxies in a LG-like environment.

In this letter, we 
present results from a simulation constrained both to form a LG
analogue and to match the velocity field of the local universe. This
allows us to contrast the properties of the simulated
LG galaxies with other galaxies of similar 
stellar mass.
For our LG simulation we have taken (a version with slightly greater coverage of) the 
ICs used to produce a MW-M31 pair 
in \citet{Nuza14} and \SCNbinprep\ in a volume
resembling the distribution of matter of our local Universe. To enlarge and contrast our sample we include the 8 galaxies of 
\citet[][S09 hereafter]{S09} which use the same code. The latter simulations are themselves 
resimulations of the 8 halos of the 
Aquarius Project \citep{Springel08} that result in isolated galaxies at 
$z=0$.

This letter is organised as follows. In Section~\ref{sec:sims} we describe the simulation code
that was used to evolve the ICs of both the LG and Aquarius, and the sample
of MW-analogues that are produced. In Section~\ref{sec:env} we analyse the evolution
of the environment of these galaxies and attempt to discern its effect on basic galaxy quantities
such as stellar mass, gas mass and star formation rate. In Section~\ref{sec:conc} we discuss and 
conclude.
 
\section{Simulations}
\label{sec:sims}

\subsection{Initial conditions}\label{sec:ICs}

The ICs used for our LG simulations 
 are part of the CLUES 
(Constrained Local UniversE Simulations\footnote{\url{http://www.clues-project.org/}}) 
Project.
The ICs 
reproduce, by construction,  the known dynamical properties
of our local environment at $z=0$ \citep{Gottloeber10,Yepes13},
and  are consistent with a \lcdm\ 
universe with WMAP-5 parameters:
$\Omega_{\rm M}=0.279$ (matter density), 
 $\Omega_{\rm \Lambda}=0.721$ (dark energy density),
 $\Omega_{\rm bar} = 0.046$ (baryon density), $H_0=100 h \,{\rm km\, s^{-1} \, Mpc^{-1}}$ with
$h=0.7$ (Hubble parameter), 
and $\sigma_8=0.8$ (normalization of the power spectrum).

The ICs use the zoom-in technique, where particles
are placed in a periodic box of
$91 \, {\rm Mpc}$ on a side
and our high-resolution region is of 10 Mpc radius at $z=0$,
with a mass resolution of $2.8\times 10^6 \, \rm M_\odot$ and  $5.6\times 10^5 \, \rm M_\odot$ 
in dark matter and gas particles respectively, and a 
gravitational softening length  of $0.7\, \rm  kpc$, 
fixed in physical coordinates  since $z=3$ and fixed in comoving coordinates
at earlier times.
The IC phases are the
same as for \citet{Nuza14}, but the zoom region is slightly larger, causing some 
variation in the simulated evolution.

The Aquarius simulations that we use to compare with our LG
galaxies are the eight galaxies first presented in S09.  These 
are the hydrodynamical counterparts of the galaxies of the  Aquarius
Project,
selected to have formed in isolated environments by requiring that
they have
no neighbour exceeding half their mass within a sphere
of $1.4\, \rm Mpc$ at $z=0$.
The cosmological parameters are slightly 
different to those used in our LG simulation: 
$\Omega_{\rm M}=0.25$, 
 $\Omega_{\rm \Lambda}=0.75$,
 $\Omega_{\rm bar} = 0.04$, $h=0.73$, and $\sigma_8=0.9$. 
The mass resolution  and softening lengths adopted are however similar in the 
two samples,
and we have used the same set of input parameters for star formation and feedback.
The different cosmological parameters used
are not expected to compromise our results as the adjustment to the cosmology is rather minor
in comparison with the dispersion in the evolutionary trends.

\subsection{Simulation code}

We use an extended version of the Tree-PM Smoothed Particle Hydrodynamics (SPH) 
code {\sc gadget3} \citep{Springel08} that includes
metal-dependent cooling, chemical enrichment and feedback
from Type II and Ia supernovae (SNe), 
a multiphase gas model and a UV background
field \citep{HM96}. The model has been developed 
in \cite{S05,S06} and is the same code as for \cite{Nuza14}.

In previous work, we have shown that our model is able to
reproduce the formation of galaxy discs from cosmological
ICs \citep{S08,S09} and alleviates
 the angular momentum problem. Some limitations 
exist, such as simulated galaxies tend to have overly massive
bulges 
(though see the implementation of \citealp{Aumer13}
 where the bulge mass is significantly
reduced via early stellar feedback), however
 discs do have
realistic sizes and angular momentum content
(\citealt{S08, S09, S10, S12}), allowing
studies of their formation and evolution in
relation to those structures on larger scales.  
In a companion paper (\SCNinprep) we analyse the 
evolution of the distribution of the stellar component,
to which we refer the reader for details of morphology
classification.

\subsection{The Galaxy Sample}\label{sec:sample}

\begin{figure}
\begin{center}
\includegraphics[width=\columnwidth]{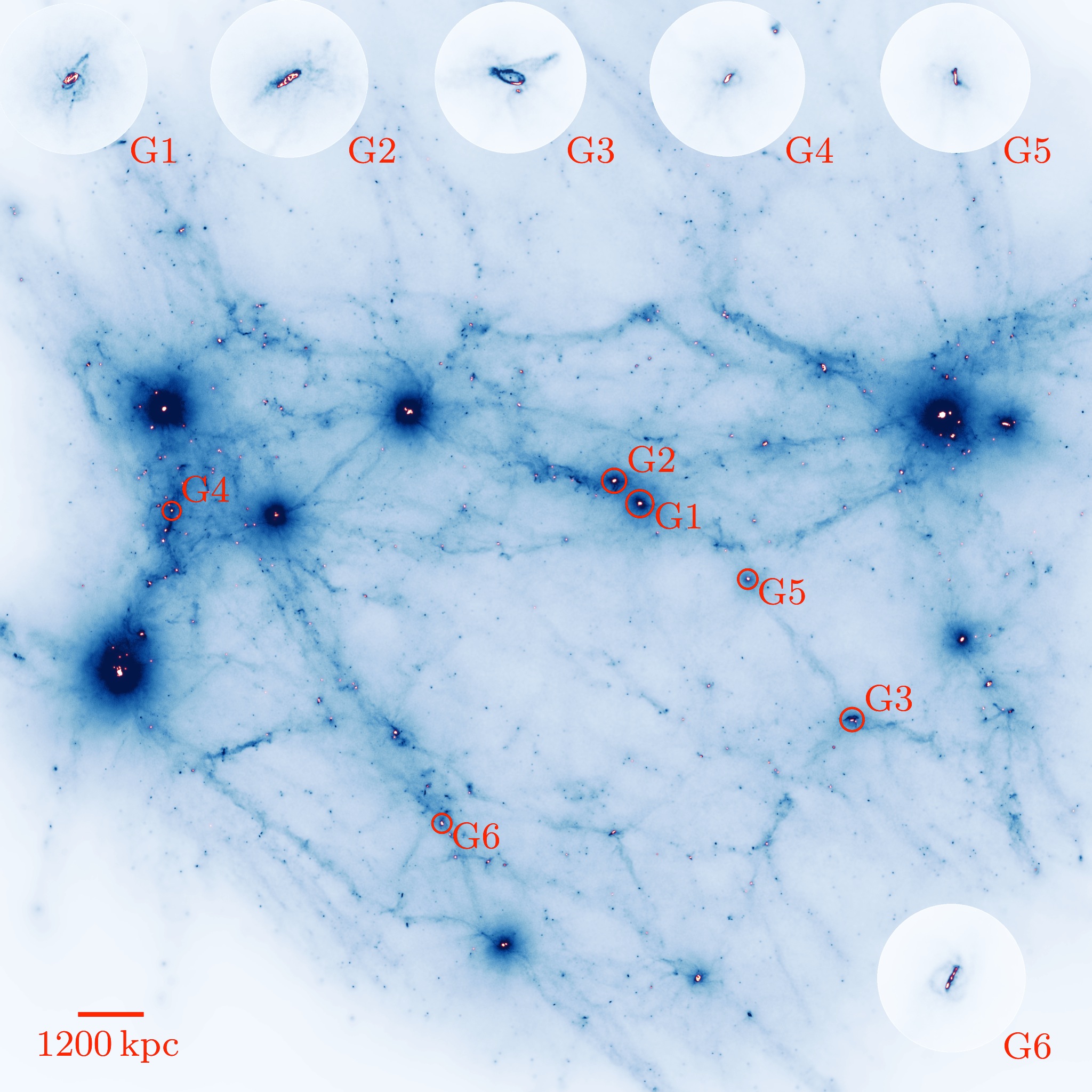}
\caption{Column density of gas (\emph{blue}) overlaid with star forming gas (\emph{orange-white}) 
for the zoom region at $z=0$. \emph{Red circles} delimit the $\Rvir$ 
for each galaxy of a similar stellar mass to the MW (G1-G6), and the labelled insets enlarge these regions.
$1200$~kpc is the radius of our environment measure, and the width of the image is $20\, {\rm Mpc}$.}
\label{fig:gas_sf_image}
\end{center}
\end{figure}

\begin{table} 
\begin{small}
\caption{ 
Main properties of the LG 
and  Aquarius galaxies at $z=0$:
  virial radius ($\Rvir$),
 virial mass ($\Mvir$), and masses in  stars, gas and star-forming gas
($\Mstar$,
   $\Mgas$ and $\MSF$), within the virial radius.}
\label{table:aquarius}
\begin{center}
\begin{tabular}{lcccccc}
\hline
   & $\Rvir$  & $\Mvir$ & 
 $\Mstar$&$\Mgas$  &$\MSF$\\
        &  {\tiny [kpc]}& {\tiny [$10^{10}\, \rm M_\odot$]} & { \tiny[$10^{10} \, \rm M_\odot$]} & {\tiny [$10^{10} \, \rm M_\odot$]} &{\tiny [$10^{10} \, \rm M_\odot$]}\\
\hline
G1 & 245 & 168 &  8.1 & 6.6 &  0.479 \\
G2 & 219 & 120 & 6.4  & 5.2 & 0.369 \\
G3 & 211 & 108 & 6.8 & 3.6 & 0.078 \\
G4 & 166 & 52 & 3.3 & 1.8 & 0.146 \\
G5 & 177 &  63 &  4.4 & 1.8 & 0.151 \\
G6 & 174 &  60 & 3.7 &  1.5 & 0.090 \\
\hline

Aq-A   &   232   &   149 &    9.2 &  4.9 &  0.153\\
Aq-B    &  181 &      71 &    4.0 &  1.7 &  0.033\\
Aq-C    &  237 &      161 &    11.0&    3.8& 0.119\\
Aq-D&      233 &     149 &     8.4&   3.5&  0.005 \\
Aq-E   & 206 &      108&     8.4&   2.6&   0.048  \\
Aq-F &   196  &   91    &    7.7 & 1.8   & 0.012  \\
Aq-G     &    180   &   68& 4.5 & 1.6&  0.061    \\
Aq-H     &   182&      74&   6.5 &  0.6  &   0.011\\

\hline
\end{tabular}
\end{center}
\end{small}
\end{table}

We choose a sample of galaxies with stellar masses in the range 
 $10^{10.5 {\rm -}11.5} \, {\rm M}_\odot$, i.e. similar to the MW.
For our LG simulation this includes the M31 and MW 
candidates\footnote{Note that in \cite{Nuza14} these galaxies were referred to as M31$^{\rm c}$ and MW$^{\rm c}$ respectively.}
G1 and G2 along with four other galaxies that
we refer to as G3-G6. These galaxies have no `contamination' within 1200 comoving kpc (ckpc) from their centres, i.e. they are sufficiently 
embedded in the zoom region to avoid low-resolution particles. 
The environment of these galaxies in gas and star-forming gas is projected in Fig.~\ref{fig:gas_sf_image},
along with insets to enlarge the properties within the virial radius. All the galaxies
in the LG sample are well separated ($>3 \,\rm Mpc$), except the G1-G2 pair 
at $770~\rm kpc$.\footnote{A visualisation of a revolution of this structure and the environment selections
is available \href{http://www.aip.de/en/research/research-area-ea/research-groups-and-projects/cosmology/galaxy-formation}{\bf here}.}

The eight Aquarius galaxies (denoted AqA-H) inhabit disjoint environments and were resimulated individually.
Images of these galaxies can be seen in S09 and for previous studies of their
formation histories and properties
we  refer the reader to \cite{S09,S10,S11}. 
Table~\ref{table:aquarius} summarises the main properties
of the LG and  Aquarius galaxies at $z=0$ inside
their respective virial radii $\Rvir$, where
the density is $200$ times the
critical density $\rho_{\rm c}$ at the corresponding redshift.

The LG
galaxies have stellar components that all show some measure of 
rotational support, however when we examine the components of angular momenta
(e.g. as in S09) we see that only half, G2, G3 and G4, have a significant fraction $(>17\%)$ of the stellar mass in a rotationally-supported disk. Nevertheless,
all these galaxies  exhibit extended gas discs and star formation at redshift zero. In the 
case of the Aquarius simulations each galaxy was able to grow extended discs during their 
evolution, but again only half can survive until $z=0$,  Aq-C, Aq-D, Aq-E and Aq-G. In both
cases we find the survival or destruction of discs in the S09 sample
was found to depend primarily on the occurrence of major mergers and 
the alignment between the angular momenta of the stars
and gas in the inner regions (see also 
\SCNinprep).

\section{Environmental effects}\label{sec:env}

The constrained nature of our simulation allows us to explore
the possibility that environment plays a role in the determination
of the properties of galaxies like our MW.
In this section, we investigate this by comparing
the properties of the LG and Aquarius galaxies.
We compare properties that are expected to depend on environment 
(gas mass, stellar mass, star formation rate), to a measure of the environment.
We note that the number of galaxies in our samples is too small to determine quantitative trends,
but that systematic differences may still be visible within the sample.

For our environmental measure we use the ratio of the mean density of
\emph{matter} within 1200 ckpc to the mean density of matter of the universe, i.e.
\begin{equation}\label{eq:env}
\denv \equiv \frac{  \left< \rho_{\rm M} \right>_{r<1200 \, \rm ckpc} }{\Omega_{\rm M}(z) \rho_{\rm c}(z)}  \, ,
\end{equation}
To give some physical reference, this radius
corresponds to how far a baryon
travelling 
at $100\, \rm km\,s^{-1}$
 traverses in $12\, \rm Gyr$ (ignoring cosmological expansion), approximately identifying the cosmological
neighbourhood which can affect a MW-mass halo. 
This scale was also used as proxy for the `local volume' by
\citet{GarrisonKimmel_2014}, and in the case
of the LG 
simulation is large enough to include
the halos of both G1 and G2 at the present time, but for all other 
galaxies forms disjoint volumes. We have considered scales from $600$-$1500\, \rm ckpc$
and found this to give the least stochastic results.

\begin{figure}
\begin{center}
\includegraphics[width=\columnwidth]{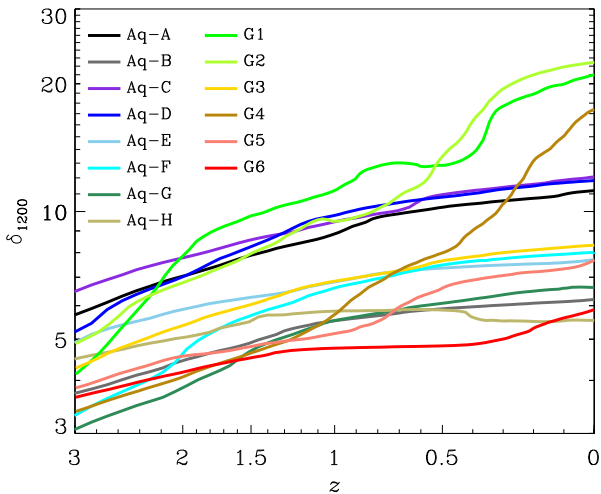}
\caption{Environmental overdensity $\denv$ vs. redshift 
for the galaxies in the LG and Aquarius samples.}
\label{fig:overdens}
\end{center}
\end{figure}

Fig.~\ref{fig:overdens} compares the environments of our 
galaxies as a function of redshift, giving us a measure of the
environmental `assembly', although we would stress that at these large 
radii we are not seeing halo mergers but rather a growth in the 
richness of the environment. Apparent is that although all the galaxies inhabit similar environments at 
$z\sim 2$-$3$, the three galaxies G1, G2 and G4 exhibit stronger 
evolution after $z\approx 2$ and are the galaxies with the 
highest overdensity $\denv$ at 
$z=0$, approximately twice as overdense as the Aquarius 
galaxies and the remaining LG galaxies. 
Additionally these three all inhabit gaseous filaments, the former two reside in the filament identified by \citet{Nuza14} and for the latter can be seen in the animated (online) version of Fig.~\ref{fig:gas_sf_image}.  Such structures have previously been implicated as affecting SF in groups \citep{Bahe_2013}.
Since these three galaxies 
recur as outliers later we denote them the `rich sample' and by contrast
the remaining 11 as the `poor sample'. 
We note that this evolution of $\denv$ implies
imprints due to environment on the rich sample 
will disappear at early epochs ($z\gtrsim 1$), before they were outliers.

\begin{figure}
\begin{center}
\includegraphics[width=\columnwidth]{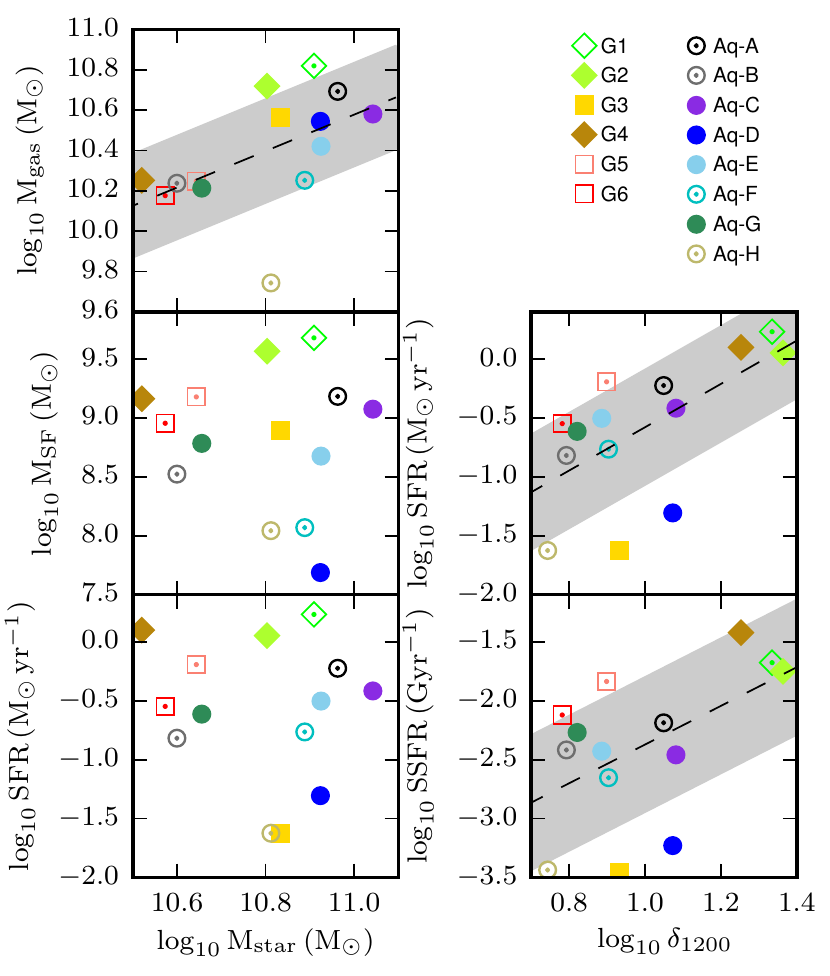}
\caption{Comparison between the properties of the galaxy samples at $z=0$. 
\emph{Left column} from top to bottom shows the gas mass ($\Mgas$), the star forming gas mass ($\MSF$) and the star formation rate (SFR), all as a function of stellar mass  ($\Mstar$). \emph{Right column} shows the SFR and the specific star formation rate (SSFR) with $\denv$ respectively. \emph{Filled symbols} indicate the galaxies that have stellar disc, \emph{diamonds} are those in the rich sample and \emph{grey-shaded regions}, where present, indicate the significant linear regressions with $\pm 1\sigma$ errors.}
\label{fig:aq_lg}
\end{center}
\end{figure}

In the left hand column of Fig.~\ref{fig:aq_lg} we compare the distribution of 
three fundamental properties of the galaxies at $z=0$:
the gas mass ($\Mgas$) and the star forming gas mass ($\MSF$)
within $\Rvir$, and the star formation rate (SFR) for material 
within $30\, \rm kpc$ averaged over the last $500~\rm Myr$.
These quantities are all plotted as
a function of the present-day stellar masses ($\Mstar$) and we additionally
mark the galaxies with stellar discs.

The cosmological trend of $\Mgas$ to rise with $\Mstar$ is visible even 
within this restricted stellar mass sample, and indeed the significance of the trend
shown is $4.4\sigma$. For $\MSF$ and SFR the trends are 
insignificant, partly due to the higher stochasticity
of these quantities (note the dynamic range of $\MSF$ and the SFR is much
higher than just for the gas) that obscures the trend in such a
limited sample, both in terms of small number and restricted stellar mass 
range. More interestingly, perhaps, is that two of the rich sample (G1 and G2)
 are the outliers in $\Mgas$, $\MSF$ and SFR, and if we only consider SFR then the
entire rich sample is extremal. 
This suggests 
the possibility that the richer environments of these galaxies are contributing
to higher star formation rates than those found in the poor sample. There does
not appear to be corresponding trends with the morphology (i.e. whether the stellar component
lies in a disc)
as was also noted by \citet{Few12}, and indeed in \SCNbinprep\  
we find morphology to be largely determined by the merger history. 

These hints of an effect of the denser environment
are borne out by classification with environmental overdensity (Fig.~\ref{fig:aq_lg} RHS).
The SFR vs. overdensity  has significant correlation, rejecting the null hypothesis at $3.8\sigma$. 
To account for comorbidity of $\denv$ with $\Mstar$, we additionally considered
the specific star formation rate (SSFR, for which we simply use SFR$/\Mstar$) and $\Mstar$ vs. $\denv$
(not shown).
The significance of the SSFR trend is lower ($2.6\sigma$), though both 
are more significant than the correlation of $\denv$-$\Mstar$ ($0.8\sigma$),
i.e. environment is a better indicator 
of SFR than stellar mass. 
The correlation between $\Mvir$ and SFR is also poor as the virial mass closely traces $\Mstar$.

\begin{figure}
\begin{center}
\includegraphics[width=\columnwidth]{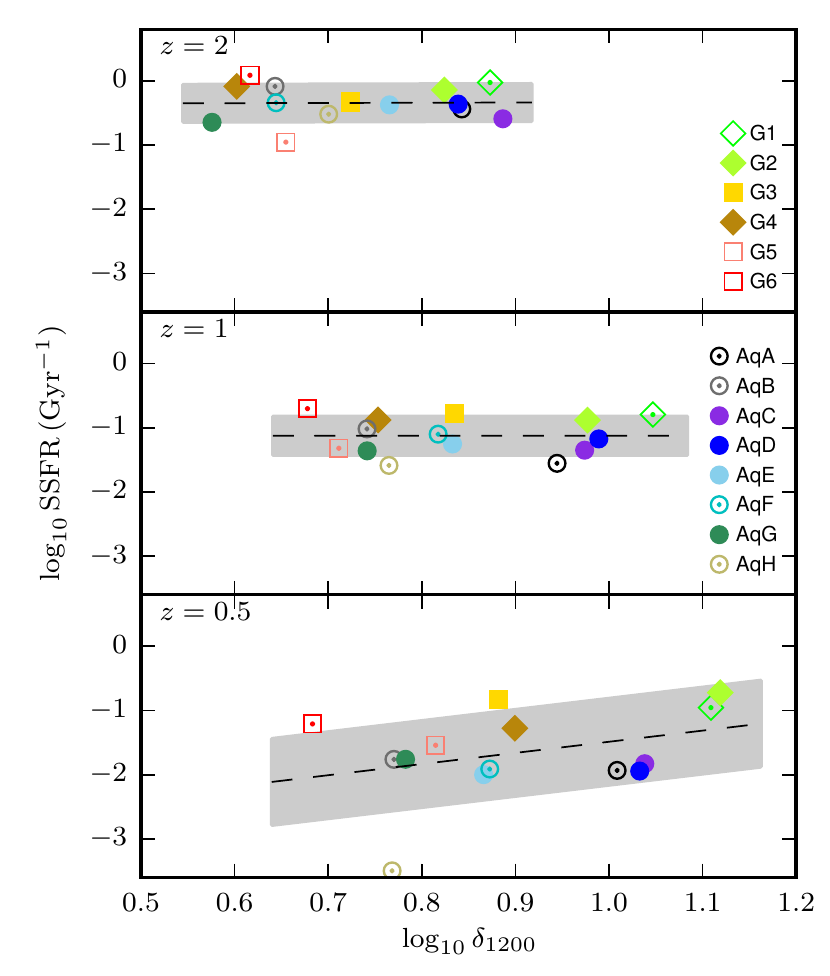}
\caption{As for the lower right panel of Fig.~\ref{fig:aq_lg} but for redshifts 
$z=2$, $1$, $0.5$
from top to bottom.}
\label{fig:Aq_LG_evolution}
\end{center}
\end{figure}

Having seen an effect of environment on SFR for MW-mass galaxies at $z=0$ it is interesting
to consider whether this exists for their progenitors at higher redshifts. Our simulations also 
provide evidence here, although we must be somewhat cautious
as galaxies are correlated with their progenitors.
In Fig.~\ref{fig:Aq_LG_evolution} we plot the
SSFR vs $\denv$ for our 14 galaxies at $z=2,\, 1$ and $0.5$. At higher 
redshifts, the galaxies live in sparser environments with higher SSFRs, and noticeably
the $\denv$-SSFR correlation fades towards earlier epochs. 

Comparing Figs.~\ref{fig:overdens}-\ref{fig:Aq_LG_evolution} we 
can see that from $z \lesssim 1$ the rich sample transitions to a denser environment
with a modest drop in SSFR whilst the poor sample has a more pronounced fall in
SSFR but less evolution in $\denv$. This suggests a possible explanation for the trends in
Fig.~\ref{fig:aq_lg}: the cold gas is driven by the richness of the 
environment, i.e. the accretion of dense structures allows the star
forming gas to replenish and higher rates of star formation to continue to
$z=0$. In our simulations 
subhalos deposit little of this due to their low gas content
 \citep{Nuza14}, although they may trigger star-formation 
indirectly.
Intriguingly, our trends for SFRs and star forming gas masses are in opposition
to that of clusters vs. the field, where suppression 
extends to group scales (e.g. \citealp{McGee_2008} though see also \citealp{Ziparo_2013}).
This may result from the density of those systems compared to our sample, 
and in the case of the LG galaxies from the fact that the LG has a late assembly, and so environmental effects will be less pronounced compared to more evolved systems.

\section{Conclusions}\label{sec:conc}

In this letter we performed
a hydrodynamical constrained simulation of the structures within
$10~\rm Mpc$ of the MW to search for Mpc-scale environmental effects
on galaxy properties. 
We compare the star formation rates, morphologies and gas fractions
of
6 galaxies of MW-like stellar masses 
to 8 from unconstrained simulations
and find systematically
higher star formation rates in the galaxies that live in richer environments, 
though this trend did not extend to morphologies.

Most notably the simulated MW and M31 candidates are overabundant both in 
present-day SSFR and in environmental density, marking them as outliers compared
to the remaining galaxies. These exceptional values disappear at earlier 
cosmological times, corresponding to their environmental assembly. The third
galaxy which displayed this trend inhabited the dense environment of a filamentary
structure.

Although the number of galaxies is small and larger
samples are needed to confirm these trends, our
results suggest that galaxies of a given mass that live in richer
environments could more easily replenish their gas
reservoirs enabling higher star formation rates.
This is consistent with a picture where the morphologies of MW-like galaxies
are primarily set by their merger history, yet environment still plays a role
in their star formation histories. This demonstrates the need
to understand in detail the effects of environment on galaxy properties 
if we are to understand
the formation of our own galaxy.

\section*{Acknowledgments}
The simulation was performed on the MareNostrum cluster in Barcelona.
CS and PC acknowledge support from the Leibniz Gemeinschaft through
grant SAW-2012-AIP-5 129. 
SEN acknowledges support from the Deutsche Forschungsgemeinschaft under the 
grants MU 1020 16-1 and NU 332/2-1, and 
GY thanks  MINECO  (Spain)  for supporting his research through different  projects:  AYA2012-31101, FPA2012-34694 and Consolider Ingenio SyeC CSD2007-0050.

\end{document}